\documentclass[preprint2]{aastex}

\begin{document}

\title{HST Observations of the Wolf-Rayet Nebula NGC 6888}
\author{Brian D. Moore, J. Jeff Hester, and Paul A. Scowen}
\affil{Department of Physics and Astronomy, \\
    Arizona State University, Box 871504, Tempe, AZ 85287-1504}

\begin{abstract}
 
We present {\it Hubble Space Telescope} WFPC2 images of a portion of
the bright northeast rim of NGC 6888, the nebular shell physically
associated with the Wolf-Rayet star HD 192163.  The exposures are taken
in the light of H$\alpha$ $\lambda 6563$, [\ion{O}{3}] $\lambda 5007$,
and [\ion{S}{2}] $\lambda \lambda 6717,6731$.  The images are used to
constrain models of the ionization structure of nebular features.  From
these models we infer physical conditions within features, and estimate
elemental abundances within the nebula.  The results of our analysis,
together with the degree of small scale inhomogeneity apparent in the
images, call into question the assumptions underlying traditional
methodologies for interpretation of nebular spectroscopy.

The thermal pressure of photoionized clumps is higher than the
inferred internal pressure of the shocked stellar wind, implying that
the current physical conditions have changed significantly over less
than a few thousand years.  These results are discussed within the
context of published three-wind evolutionary scenarios for the formation
of the nebula.  We also discuss the nature of a back-illuminated
radiative shock driven into the cavity surrounding NGC 6888.

\end{abstract}


\section{Introduction}

     Wolf-Rayet (WR) stars are thought to be the late stage
of the evolution of stars more massive than $\sim 30$ M$_{\odot}$.
WR stars are characterized as possessing significant
stellar winds, large in both mass-loss rate and terminal velocity.
Observations of 62 Galactic WR stars in the northern sky show
that about half are associated with nebular emission \citep{miller},
some of which are ring nebulae.  \cite{johnson} were the first
to propose that these ring nebulae were formed by stellar wind
interaction with the local interstellar medium.  Attempts have
been made to analytically explain the formation and structure
of such nebulae \citep{weaver,vanburen}.

One of the most prominent and best studied examples of this class of
object is NGC 6888.  NGC 6888 is associated with WR 136 (HD 192163),
considered to be a member of the Cygnus OB1 association.  Its large
angular size (18$\arcmin$ x 12$\arcmin$) has made it an attractive 
object for investigation at a variety of wavelengths.  For a
distance of 1450 pc \citep{wendker} the nebula has a physical
size of 7.6 pc $\times$ 5.0 pc.  One of the first detailed
studies of NGC 6888 was that of \cite{lozhin} who, assuming
that the nebula is prolate ellipsoid, stated that the nebula
is tilted $\sim 30^\circ$ out of the sky.  Later studies
\citep{parker, kwitter} used emission-line ratios to show
that the nebula is overabundant in nitrogen and helium and
underabundant in oxygen. Explanations for this enrichment include the
transport of processed material from the stellar interior to the outer
atmosphere prior to the onset of the RSG phase \citep{esteban}, and the
mixing of interstellar and wind material \citep{kwitter}.  The high
level of nitrogen is consistent with the classification of WR 136 as a
WN6 star.  An [\ion{O}{3}] skin external to the H$\alpha$ nebulosity
has previously been observed \citep{dufour,mitra}, suggesting that the
stellar wind and the expanding shell are interacting with the bubble
outside the RSG shell.

There have been a number of hydrodynamical simulations of the
formation of Wolf-Rayet shells like NGC~6888. The most complete
of these is \cite{garcia}, based on a model for the evolution
of a 35 M$_{\odot}$ star.  A summary of the stages in their
scenario is as follows.

During its main-sequence (MS) lifetime the star sweeps the
surrounding interstellar material into a thin shell with a radius
of tens of parsecs.  The interior of the MS bubble is filled
with hot ($> 10^{6}$ K) post-shock stellar wind material.
In the absence of thermal evaporation of material at the outer
edge, the electron density within the MS bubble should be of order
$10^{-3}$ cm$^{-3}$.  

After $4.5 \times 10^{6}$ years the star evolves off the main
sequence and into a red supergiant (RSG). Over its lifetime
of $2 \times 10^{5}$ years , the RSG wind fills the inner 
few parsecs of the MS bubble with cold dense material.
The characteristic mass loss rate of the RSG phase is
$ \sim 10^{-4}$ M$_{\odot}$ yr$^{-1}$ with terminal velocity
15 km s$^{-1}$.  Between the free-streaming RSG wind and the MS
bubble a thin dense shell of post-shock RSG material forms.

Lastly the star enters its Wolf-Rayet phase, with mass loss rate
$ \sim 10^{-4.5}$ M$_{\odot}$ yr$^{-1}$, with V$_{\infty} \sim$
2000 km s$^{-1}$ over its lifetime of $2 \times 10^{5}$ years.
The WR wind progressively sweeps up the RSG material, eventually
overtaking the outer RSG shell.  The resulting collision fractures
the outer shell and the WR bubble breaks out into the MS
bubble material.  At this point the nebular shell reaches
maximum brightness due to its high density; according to their model,
the nebula should be detectable over a period of order 10$^{4}$ years.

The morphology and dynamics of WR shells are important not only as
laboratories for studying wind interactions with the ISM, but also for
the insights they provide into the mass-loss history of massive stars
as well as the environments into which remnants of core collapse
supernovae will propagate and interact.  In this paper, we use
observations of a region of NGC 6888 obtained with the {\it Hubble
Space Telescope} WFPC2 to study the ionization structure and infer
physical conditions within the shell.

\section{Observations and Results}

\subsection{Observations}
 
     Observations of NGC 6888 were made in July 1995 using
the Wide Field/Planetary Camera 2 on the {\it Hubble Space Telescope}.
The orientation of the field of view is shown on a ground-based
image of the nebula in Figure \ref{GROUND}.  The characteristics
of the WFPC2 instrument are described in \cite{trauger}.  Three of
the cameras have an image scale of $0 \farcs 0996$ per pixel;
the fourth camera has an image scale of $0 \farcs 0455$
per pixel.  Two exposures were made in each of three filters:
F502N ([\ion{O}{3}] $\lambda 5007$), F656N (H$\alpha$), and
F673N ([\ion{S}{2}] $\lambda \lambda 6717,6731$).  The integration
time for each filter was $2 \times 1100$ seconds.  Each pair of
exposures was reduced and combined using a cosmic-ray rejection
routine.  Remaining cosmetic defects were removed by direct
inspection of the images.  The resultant CCD frames were mosaicked
using a routine utilizing the astrometric solution in \cite{holtza}.
Figure \ref{COLOR} shows a false-color composite of the narrow-band
images in a linear stretch, with [\ion{O}{3}] shown as blue, H$\alpha$
as green, and [\ion{S}{2}] as red.  Grayscale images from the
individual filters are shown in Figures \ref{HAL} (H$\alpha$),
\ref{OXY} ([\ion{O}{3}]), and  \ref{SUL} ([\ion{S}{2}]).

The conversion from DN to surface brightness (in erg cm$^{-2}$
sec$^{-1}$ pixel$^{-1}$) for the narrow-band filters using the
procedure outlined in \cite{holtzb};

\begin{equation}
       F_{\lambda} = DN \times \left( \frac{14}{GR_{i}}\right)
                   \times \frac{E_{\lambda}}{A \times
                   QT_{\lambda} \times t_{exp}}
\end{equation}

\noindent
where $DN$ = total data numbers in a given pixel,
$GR_{i}\simeq$ 2 for the gain state used (7e$^{-}$/DN),
$E_{\lambda}$ = energy of a photon at the observed wavelength, 
$A$ = collecting area of the HST, $QT_{\lambda}$ = the
telescope+filter efficiency at the observed wavelength, 
and $t_{exp}$ = the exposure time of the image.

     The flux must then be corrected for internal and interstellar 
extinction.  \cite{mitra} estimates the extinction in the NW region
from the H$\beta$/H$\alpha$ ratio in the spectra.  Results for the
three positions are C$_{H\beta}$ =  0.94, 1.05, and 1.13.
Using the median value with the reddening law given by \cite{cardelli},
the dereddened surface brightness in this region is given by

\begin{eqnarray}
       I(F656N) & = & F(F656N) \times 10^{C_{H\alpha}} 
       \nonumber  \\ & = & 5.27 \, F(F656N).
\end{eqnarray}

and similarly, 

\begin{mathletters}
\begin{eqnarray}
       I(F673N) = 5.04 \, F(F673N), \\
       I(F502N) = 10.2 \, F(F502N).
\end{eqnarray}
\end{mathletters}

\subsection{Results}

    At a distance of 1450 pc one WF pixel corresponds
to $2.18 \times 10^{15}$ cm.  The entire field of view
is thus about 1 pc on a side.  The network of clumps
comprising the nebular shell are photoionized by the
central star HD 192163, about 3 parsecs away in the 
direction indicated in Figure \ref{COLOR}.  We assume 
that the nebula is oblate; i.e., the clumps
and central star lie approximately in the plane of
the sky; hence, distances in the image do not suffer
geometric foreshortening.  Even accepting the prolate
shape of \cite{lozhin}, distances would only be
compressed by 15\%.  The photoionized clumps are visible
to varying degrees in all three filters, with a strong
correlation between [\ion{S}{2}] brightness and
H$\alpha$ brightness.  The filaments are typically
shorter along the axis pointed towards the central
star relative to the axis perpendicular to it.  The
network of filaments is enveloped by a ``skin'' of
[\ion{O}{3}] emission, formed by the shock driven
into the external material by some combination of the
nebular shell expansion and the pressure of the post-shock
WR wind.  The diffuse component of the nebular H$\alpha$ 
and [\ion{S}{2}] emission, mostly internal to this skin,
is a combination of emission from these two regions.
In the NW quadrant, one clump deviates significantly
from the positions of the rest of the shell.  It has
an [\ion{O}{3}] arc extending beyond it, as well as
a diffuse H$\alpha$ tongue of emission emanating from
the clump and extending beyond the arc itself.

\section{Photoionization Of The Nebular Shell}

     The material lost by the central star in its
previous red supergiant wind has been swept up by
the subsequent fast wind of the current WR phase.
As the shell thinned it became subject to instability
and fragmented.  The bright patches of visible
emission of NGC 6888 seen here are understood to
be the fragmented circumstellar shell photoionized
by the UV flux of the central star.

\subsection{Spatial Profiles}

     Features were chosen on the basis of high H$\alpha$
brightness.  Those clumps with significant variation
at the resolution limit were rejected.  Those clumps
that appeared to have a complicated geometry were
also rejected.   A total of six clumps across the
observed region were chosen from the remaining sample.
Spatial profiles were extracted along the direction
towards the central star with a stepsize of 0.5 pixel.
The positions of the spatial profiles are shown in Figure
\ref{CUTS}.  The background level for each profile was fit
fit by a linear or quadratic function and subtracted.
Finally, the depth along the line of sight was estimated
from the geometry of a given clump, using the geometric
mean of the major and minor axes for the spheroidal clumps
and the overall curvature of the extended clumps as a rough
equivalent to the emitting column.  The volume emissivity
of the gas is then given by

\begin{equation}
4 \pi j_{\lambda} = \frac{4 \pi I_{\lambda}}{\Omega l},
\end{equation}

\noindent
where $\Omega$ = solid angle subtended by one WF pixel,
and $l$ = estimated line-of-sight depth, in cm.

     A first estimate of the hydrogen density can be derived
from these emissivities; using \cite{oster}, Table 4.2, 
one sees that for 10$^{4}$ K gas,

\begin{equation}
4 \pi j_{\lambda} = 3.56 \times 10^{-25} 
n_{p} n_{e} {\rm \; \; erg\,cm^{-3} s^{-1}}
\end{equation}

     For the clumps chosen the surface brightness and estimated
line-of-sight depths result in densities of order 1000 cm$^{-3}$.

     The spatial profiles extracted from the images were modeled
using the photoionization code CLOUDY \citep{ferland}.  Satisfactory
fits were obtained by modeling clumps as thin slabs of constant
density with two-component linear wings to the front and rear
of the slab.  The inflection point of the wings occurs where
the density falls to one-half the density of the slab.  Thus,
each clump model has five spatial parameters: the beginning
and end of the slab,  the two inflection points, and the total
width of the clump model.  The abundances in the gas were fixed
at solar values with the exception of helium, oxygen, nitrogen,
nitrogen, and sulfur.  Helium was fixed at twice the solar
abundance, following the results of \cite{kwitter}.  Varying
this abundance did not significantly change the emissivity of
the other elements.  The other abundances were constrained
by the observed profiles.  For oxygen and sulfur, the elemental
abundance was varied to maximize agreement with the appropriate
observed profile.  For nitrogen, the abundance was chosen to
match the [\ion{N}{2}]/H$\alpha$ emission ratio to the value
oberved by \cite{parker} for the corresponding area of the
nebula.  It is necessary to include the nitrogen emission in
the model profile, as the F656N filter admits approximately the
equivalent of 15\% of the [\ion{N}{2}]$\lambda 6584$ line flux
\citep{biretta}, which for NGC~6888 can represent 20\% of the
counts in the F656N image.

     Each clump was placed at its projected distance
from the central star.  The input ionizing spectrum
was the model atmosphere for a WN5/WN6 star presented
in \cite{hillier}. For comparison to the observed 
profiles the model profiles were convolved with a
gaussian of width corresponding to the PSF in the
images, approximately 0.5 pixel.

\subsection{Model Results}

     A representative comparison of an observed and model profile
is shown in Figure \ref{MODEL}.  Models with terminating
ionization fronts consistently overestimated the separation of
the peak [\ion{S}{2}] and H$\alpha$ emissivities, suggesting that
the shell does not contain large amounts of neutral material.
This is supported by the absence of [\ion{O}{1}] in the 
spectrophotometry of \cite{mitra}; based on observed line fluxes,
I([\ion{O}{1}]$\lambda 6300$) $<$ 0.02 I(H$\beta$).  The best
fit photoionization models required a minumum H-ionizing flux
of log$Q_0$ = 49.2, with significant deviations for two of the
models (clumps 2 and 5) for values less than 49.3.  This is in
reasonable agreement with the value of 49.0 from the Wolf-Rayet
models of \cite{crowther}.  It is, however, at odds with
\cite{mandm}, who calculate the Lyman continuum flux required
to maintain the observed H$\alpha$ brightness as log$Q_0$ = 47.6.
The implications of this discrepancy are addressed in \S 4.

Models were constructed for log$Q_0$ from 49.2 to 49.4.  The
uncertainty on the abundance for an individual model is small;
variation in the model elemental abundances by as little as
0.02 dex was sufficient to shift the emissivity of the corresponding
line out of agreement with the observation.  Using the traditional
nomenclature the abundances are (12 + log(N/H)) = $8.1 \pm 0.2$,
(12 + log(O/H)) = $7.8\pm 0.2$, and (12 + log(S/H)) = $6.8 \pm 0.2$.
The quoted uncertainty represents the spread between the abundances
of all of the models, summarized in Table 1 and shown graphically
in Figure \ref{ABUN}.

One of the primary sources of error in determining abundances with
this methodology is the uncertainty in the estimated line-of-sight
depth for a given clump.  This somewhat counterintuitive result
arises from the fact that, while the ionic abundance can be
inferred directly from the observations independent of density
or line-of-sight depth, the ionic fraction differs significantly
with changes in the density and ionizing flux (i.e., changes
in the ionization parameter).  For example, a 25\% error in the
estimated line-of-sight depth is roughly equivalent, in its
effect on abundances, to an error of about $\pm 0.05$ dex in
$Q_{0}$. This typically corresponds to an uncertainty of about
$\pm 0.1$ dex in abundances.  In addition, the offset between
the peak [\ion{O}{3}] and H$\alpha$ emissivities were
consistently underestimated by the models, which we ascribe to
the simplicity of the one-dimensional profile model.  This
deviation suggests that our oxygen abundance may be systematically
low; based on the model electron density at the position of the
observed peak, this effect is at most 0.15 dex.

\subsection{Comparison with Ground-Based Spectroscopy}

A longstanding question in nebular analysis involves the degree
to which spectroscopy integrates over structure within nebulae,
and the effect of inhomogeneity on inferences based on those data
(e.g., the $t^2$ parameter in Peimbert 1967).  Indeed, our images show
significant density and ionization structure on spatial scales of
$1\arcsec$ and shorter.  A single resolution element typical of ground
based spectroscopy would necessarily sample a range of physically
diverse regions.  Further, discrete features are typically embedded
within diffuse line-of-sight background that is comparably bright to
the features themselves.  It is unreasonable to imagine that
ensemble-average conditions measured spectroscopically have much
physical meaning for specific locations within the nebula.  Nor,
conversely, can regions sampled spectroscopically be described
meaningfully using a single set of physical parameters such as
temperature or density.

An examination of inferred electron densities makes the point.
Spectroscopy \citep{kwitter, mitra} of this region typically
yields [O~II] and [S~II] electron densities in the range from
100 to 450 cm$^{-3}$, well below our densities of 1000-1600
cm$^{-3}$.  As discussed above, our density estimates are based
on no more than the physics of hydrogen recombination.  Enforcing
spectroscopically-determined densities on features herein would require
that their line-of-sight depths exceed their projected sizes by one to
two orders of magnitude.  A simpler explanation is that ground-based
spectroscopy samples not only discrete features, but also unavoidably
includes large volumes of lower density, more diffuse material.
Roughly speaking, a mix of equal amounts of [\ion{S}{2}] emission from
gas at 1500 cm$^{-3}$ (typical of our inferred densities) and
[\ion{S}{2}] emission from gas in the low density limit will yield a
``measured'' spectroscopic density of around 500 cm$^{-3}$ -- {\it even
though there is little or no gas with that density actually present
within the sampled volume.}  Comparable density inhomogeneities also
exist within discrete features themselves.  Similarly, our models show
that T$_e$ can easily vary by 1000 K within the region sampled by a
spectrograph aperture, even in the absence of background emission.

To employ traditional methods of interpreting spectra (e.g., i$_{CF}$s)
it is necessary to assume a certain degree of homogeneity within
a zone responsible for emission from a given ion, as well as some
correspondence between zones responsible for emission from different
ions.  These {\it ad hoc} methods have a definite advantage -- they
yield an answer.  However, the physical assumptions upon which
they are based are not well supported by our data or our models.  While
we do not observe the wealth of lines present in a spectrum, we are able
to use a few lines from both low and high ionization species to constrain
realistic models of the ionization structure of observed features.
Our resulting abundance determinations differ from the results of
\cite{kwitter} by +0.3 dex, -0.7 dex, and and +0.5 dex for N, O, and S,
respectively.  The concerns that we raise are in addition to those of
\cite{alex}, who found that the i$_{CF}$ methodology may fail even in
cases where the density is homogeneous.  While the method we employ
embodies its own suite of uncertainties, we suggest that there is ample
evidence to warrant increased skepticism regarding the results of
traditional nebular abundance determinations.

\section{Nature of the Skin Emission}

The dense shell is enveloped in a thin skin of emission, best seen at
the outer edge of the nebula beyond the network of clumps, and most
evident in [\ion{O}{3}].  A perpendicular cut taken into the nebula at
a bright part of the skin is shown in Figure \ref{SHOCK}.  Notice that
there is a rise in [\ion{O}{3}] emission, followed by a rise in H$\alpha$
a few $10^{16}$ cm closer towards the star.  If the difference in
ionization state across the shell were due to photoionization effects,
we would expect that the higher ionization emission would be found interior
to the lower ionization emission, as seen in the clump profiles discussed
above.  Rather, this morphology suggests instead that the leading edge
of [\ion{O}{3}] emission arises in the cooling behind a radiative shock
driven into the relatively low density medium surrounding the shell.
The recombination region, however, is ``incomplete'', and the postshock 
flow cools and recombines only to the point that it comes into
photoionization equilibrium with the ionizing flux from the star.
Since the ionization parameter is high in this trailing photoionized
zone, the [\ion{O}{3}]/H$\alpha$ ratio remains high ($\sim$ 5) and 
[\ion{S}{2}] remains very weak throughout.  There are, however, 
a few positions where post-shock material is more prominent in H$\alpha$.
These are probably locations where the shock is partially ``shadowed''
from the stellar UV flux by dense clumps within the shell.

An arc seen in the NW quadrant of the field in [\ion{O}{3}] emission
extends well beyond the general perimeter of the skin.  This is
morphologically consistent with the early stage of a ``blowout'':
a place where the shocked stellar wind has broken more completely
through the nebular shell material and is expanding outward into
the MS bubble itself.  An echelle spectrum of this arc was taken
in the light of [\ion{O}{3}] in December 1997 by \cite{graham}
using the 3.5-m telescope at Mt. Hamilton.  We do not discuss
this spectrum in detail here other than to note that it showed the
characteristic velocity structure of an expanding bubble at this
location.  Assuming the line-of-sight angle of the expansion is the
same as the angle of the arc to the general perimeter of the skin,
the echelle arc has in an expansion velocity of $70 \pm 12$ km s$^{-1}$,
implying a shock velocity of $\leq$ $93 \pm 16$ km-s$^{-1}$.  This
is insufficient to produce such a large [\ion{O}{3}]/H$\alpha$ in
a neutral pre-shock medium, suggesting that the pre-shock medium
is fully ionized \cite{cox}.  We show below that this is supported
by observation.

     We estimate the emissivity of the post-shock
gas at the leading edge of the NW arc, again using
the geometry to estimate the pathlength.  Assuming
a temperature of $10^{4}$ K, this yields a hydrogen 
density of 100 cm$^{-3}$.  Due to Poisson noise,
the pathlength estimate, and the unknown temperature,
the uncertainty is about a factor of 2.  If we assume
the pre-shock gas temperature is also $10^{4}$ K,
then we can use the isothermal assumption to relate
the densities of the pre-shock MS material and post-shock
skin material; i.e.,

\begin{equation}
     \rho_{skin} = 2 M^{2} \rho_{MS}.
\end{equation}

\noindent
With the shock velocity 93 km s$^{-1}$, M = 5.6,
and the resulting pre-shock density is 2 cm$^{-3}$.

On the basis of our analysis, it is clear that the shell is
extremely ``leaky'' to ionizing emission.  Using parameters
derived below, the optical depth of the postshock region to
ionizing radiation is only about $2 \times 10^{-3}$ assuming
log$Q_0$ = 49.3, and so any radiation that reaches the shock
has effectively escaped from the nebula.   The densest knots
themselves have optical depths of order 1.  As can be seen
from the more face-on portions of the shell in \ref{GROUND},
such dense knots cover a very small fraction of the surface
of the nebula.  As a result, Zanstra methods cannot be used
to obtain reliable estimates of the total H-ionizing flux of
the central stars of objects such as NGC~6888.  Based on the
H-ionizing flux budget of the nebula obtained by \cite{mandm}
(log$Q_0$ = 47.6) and the value derived in this work (49.3),
{\it only 2\% of the ionizing photons from HD 192163 are
processed within the nebular shell of NGC 6888.}

Another consequence of this analysis is that any material
surrounding NGC 6888 will be fully ionized.  This implies
that there cannot be a significant amount of neutral material
close to the observed nebula as suggested by \cite{marston}.
However, an ionized shell of density 2 cm$^{-3}$ exterior to
the nebular shell could contain a significant amount of mass
with low surface brightness.  It may be that a major portion
of the estimated 18 M$_{\odot}$ of material lost in the RSG
wind has been lost to the MS bubble, and exists as a low
density shell is only visible in post-shock emission.

\section{Discussion}

     NGC 6888 is primarily composed of the material from
the RSG wind of the Wolf-Rayet progenitor, swept into 
a shell by the central star's WR wind.  This shell 
is photoionized by the UV flux from the central star.
We combine the results of this paper and previous
X-ray observations to create a coherent picture of the
past and present state of NGC 6888.  In Figure \ref{SKETCH}
is a sketch of the optical and X-ray emitting regions.

\subsection{Gas Pressure}

     The gas pressure of a shocked stellar wind can
be expressed in terms of its mass loss rate and its
terminal velocity.  For an adiabatic shock, the 
post-shock gas pressure is

\begin{equation}
P_{w} = \frac{3 \dot{M} V_{\infty}} {16 \pi R_{w}^{2}}
             = 2.88 \times 10^{7} \dot{M}_{-4}
             V_{3} R_{w}^{-2} \, {\rm cm^{-3} \, K},
\end{equation}

\noindent
where $\dot{M}_{-4}$ is the mass loss rate in
units of $10^{-4}$ M$_{\odot}$ yr$^{-1}$, $V_{3}$ is
the terminal velocity in units of $10^{3}$ km s$^{-1}$,
and $R_{w}$ is the radius of the termination shock
in parsecs.
 
     \cite{crowther} derive the stellar parameters
$\dot{M}_{-4} = 1.23$ and $V_{3} = 1.75$ from model
fits to infrared observations of WR 136.  In addition,
they also summarize the results of previous studies
for this object.  If we assume that the inner edge
of the X-ray emission corresponds to the location
of the termination shock, we estimate that $R_{w} = 3.1$ pc.
These results yield a pressure behind the stellar
wind shock of

\begin{equation}
P_{w} =  3.4 - 5.3 \times 10^{6} \, {\rm cm^{-3} \, K}.
\end{equation}

\noindent
assuming that the location of the termination
shock is not changing rapidly.

     The X-ray flux comes from the post-shock WR wind
material.  For the X-ray emitting region, the results
of \cite{wrigge} were used for the temperature, emission
measure, and emitting volume.  (We note that we suspect 
the value quoted in their paper for the emission measure
has a transcription error, resulting in an error of a
factor of $10^{28}$.  We use this corrected value in this
paper.)  The density can be derived from the emission
measure E; i.e.,

\begin{equation}
n_{H}^{2} V \sim E \times 4 \pi D^{2},
\end{equation}

\noindent
where V is the volume of the emitting region, and 
D is the distance to the nebula.  \cite{wrigge} get an  
emitting volume of $7-12 \times 10^{55}$ cm$^{3}$ and
a (corrected) emission measure of $(1.5 \pm 1.1) \times
10^{11}$ cm$^{-5}$, then at an assumed distance of 1450
pc the density of the X-ray gas is n$_{H}$ = 0.3 - 0.8
cm$^{-3}$.  With a temperature of $2.36 \pm 0.52 \times
10^{6}$ K, the pressure of the X-ray gas is

\begin{equation}
P_{X} =  1.4 - 3.8 \times 10^{6} \, {\rm cm^{-3} \, K},
\end{equation}

\noindent
roughly equal to that of the post-shock stellar wind.
This gas is not simply the post-shock wind material,
as its expected temperature is $4.2 \times 10^{7}$ K
and of such low density that it would be unable to
cool quickly to the observed X-ray gas temperature.
Instead, this material is probably the result of thermal
evaporation of nebular shell material into the post-shock
stellar wind \citep{hartquist}.  The rough equivalence
between the pressure expected from eq. (8) and the
pressure of the observed gas from eq. (10) supports
this contention.

     The gas pressure of each of the regions can be
calculated assuming an ideal gas law.  For the outer
shock and nebular shell, the values used are those
derived in the previous sections.  For the outer shock,
the pressure is 

\begin{equation}
P_{skin} \sim  2 \times 10^{6} \, {\rm cm^{-3} \, K}.
\end{equation}

\noindent
Given uncertainties in parameters used to infer pressures for the
various components of NGC 6888, this is not significantly different
from the pressures inferred for
the shocked stellar wind and the X-ray gas.  

     For the clumps comprising the nebular shell
the central pressure can be calculated from the
models.  Using a peak density of 1300 cm$^{-3}$
at a temperature $10^{4}$ K typical of the spatial
profile models yields a pressure of

\begin{equation}
P_{shell} =  26 \times 10^{6} \, {\rm cm^{-3} \, K}.
\end{equation}

\noindent
Thus, clumps in the nebular shell are in severe 
overpressure compared to the surrounding gas, by
a factor of order 10.  This is consistent with the 
inclusion of density ramps in the model profiles of
these regions, although the magnitude of the discrepancy 
is surprising.  With typical clump size of $10^{17}$
cm, isothermal expansion of the clumps at the sound
speed should alleviate this pressure discrepancy
on a timescale of order 2000 years.  This implies  
that until recently the termination shock was at
R$_{w} < 1.3$ pc.  If we assume that the loss of
pressure support for the termination shock would
result in its motion outward at the stellar wind
velocity, it would take of order 1200 years for
R$_{w}$ to move from 1.2 to 3.1 pc.  Considering
that the clump pressure was initially higher than
now observed coupled with the quadratic dependence
of pressure with R$_{w}$, there can be a 
considerable time where the clumps have not yet 
come into pressure balance with its surroundings.

Is there evidence supporting the contention of
rapid depressurization of the bubble interior?
Initially, we note that the thickness of the
postshock region behind the outer [\ion{O}{3}] skin
provides an estimate of the age of this shock.
Assuming a post-shock density of 100 cm$^{-3}$,
a preshock density of 2 cm$^{-3}$, a shock velocity
of $\sim$ 100 km s$^{-1}$, and a thickness of
$10^{16}$ cm, the shock broke through the densest
portion of the shell about 1500 years ago.
Blowouts can also be seen on the global scale of
NGC 6888.  In \cite{miller} Figure 3b, one sees
that the elliptical nebular shell is enclosed
by a complete, approximately spherical
[\ion{O}{3}] bubble.  The loss of containment
would cause a rapid drop of internal pressure
like that implied by the pressure mismatch between
the shell and stellar wind.  If we use the
time for the clumps to expand into balance with
the new pressure as a constraint, then the edge
of the blowout has to traverse 1.3 pc in much less
than 2000 years, requiring an expansion velocity
well over 600 km s$^{-1}$.  This implies that
the density in the outer MS bubble drops with
increasing distance.  Using the present internal
pressure as a lower limit, the density of the
outer MS bubble is still $> 0.1$ cm$^{-3}$.

\subsection{Comparison with Hydrodynamical Simulations}

     The most complete numerical simulations of 
Wolf-Rayet bubble formation and dynamics is
\cite{garcia}.  In it, they follow the evolutionary
phases of a massive star, with main-sequence, red
supergiant, and Wolf-Rayet mass loss.  In summary,
they produce an MS bubble 30 pc in radius, filled
with hot ($10^{7}$ K) thin (0.002 cm$^{-3}$) material;
18.5 M$_{\odot}$ of RSG material, with a thin shell
of post-shock RSG material at the RSG-MS interface;
and a Wolf-Rayet wind that sweeps up the RSG material,
leading to a collision with the thin RSG shell and
eventually breakout of the WR wind into the MS bubble.
The results of this paper show two main deviations
from this picture.  First, as argued before, the 
material filling the MS bubble is at least 0.1 cm$^{-3}$.
For the MS bubble density implied by \cite{garcia},
any outer shock would not be visible.  The thermal pressure 
of the observed material is the same as that assumed for
the simulations, so a thin RSG shell can still form.
Second, their post-collision shell density (1000 cm$^{-3}$)
is consistent with that derived in this paper, but the total
mass of the shell is 4 times higher than that implied 
by our result of a fully ionized shell.  A simple
resolution to both of these differences is the
presence of thermal evaporation from the thin RSG
shell into the MS bubble.  At 2 cm$^{-3}$, a $\sim$ 1 pc
thick annulus of material outside the nebular shell would
have a mass of 15 M$_{\odot}$ of material and still be 
undetectable in the WFPC2 image.  A smaller annulus with
a density gradient could hide a similar mass.  It has also 
been suggested that the central star's proper motion
can remove it from its relic MS bubble; this does
not resolve the RSG mass loss issue.  Another
minor issue is that of the limitation of these
simulations.  In a non-spherical shell the pressure
will drastically change before the shells collide
along the major axis.  How that alters the dynamics
of the shell is undetermined, although the problem
of ellipsoidal shells was pursued in \cite{garciaold}.

\section{Conclusions}

     NGC 6888 is a bubble formed by the mass loss of
the precursor to the central star WR 136.  The internal
pressure of the shocked WR stellar wind has swept up
the ejecta from the previous RSG phase.  This material 
collides with a thin RSG shell formed at the MS bubble
interface, fragmenting into the clumps seen today.  Models
of spatial profiles of selected clumps appear to be fully
ionized, resulting in a log Q$_{0} >$ 49.2 for the central
star, and average abundances for the shell material of
[N, O, S] = [8.1, 7.8, 6.8].  The oxygen depletion
appears even more severe than that seen in previous
spectroscopic studies.

     The internal pressure of the stellar wind drives
a shock into the material external to the visible shell.
The shock is visible in [\ion{O}{3}] and H$\alpha$ as a
skin enveloping the clump network.  The density of this
material suggests that the MS bubble cooled efficiently.
Combined with the mass of the visible shell compared to
the total mass loss of the RSG phase, we argue that
thermal evaporation of material from the outer RSG shell
has already occurred.  Comparing the pressure inside
and outside the nebular shell suggests that a radical
drop in pressure has recently occurred, and that the
clumps have not yet come into equilibrium with the
current pressure of the stellar wind.

\section{Acknowledgments}

     BDM would like to thank Jason P. Aufdenberg and Ravi
Sankrit for helpful comments in the preparation of this paper.
The authors would also like to thank Dr.~Hillier for providing
his WN5/WN6 model atmosphere for the preparation of this paper.
This work was supported by NASA grant NAS 5-1661 to the WF/PC
Investigation Definition Team (IDT) and NASA contract NAS-7-1260
to the WFPC2 IDT. This work was supported at Arizona State
University by NASA/JPL contracts 959289 and 959329 and Caltech
contract PC 064528.

\newpage

\begin{figure}
\caption{Ground-based image of NGC 6888, showing the field
    observed with WFPC2.  North is at the top and east is to
    the left.  The WFPC2 footprint is approximately $150 \arcsec$
    (1 pc at a distance of 1450 pc) on a side.}
\label{GROUND}
\end{figure}

\begin{figure}
\caption{False-color composite of the WFPC2 images of NGC 6888.
    Emission from [\ion{O}{3}] is shown in blue.  Emission from H$\alpha$
    is shown in green.  Emission from [\ion{S}{2}] is shown in red.}
\label{COLOR}
\end{figure}

\begin{figure}
\caption{WFPC2 image of NE section of NGC 6888 taken through
    the F656N filter, which isolates light from H$\alpha$ $\lambda
    6563$.  North and east are indicated by the arrows.  The bar
    has a length of $10 \arcsec$, corresponding to $2.2 \times
    10^{17}$ cm.  The negative display uses a square-root stretch
    to allow comparison over a large range in surface brightness.}
\label{HAL}
\end{figure}

\begin{figure}
\caption{WFPC2 image of NE section of NGC 6888 taken through
     the F502N filter, which isolates light from [\ion{O}{3}] $\lambda
     5007$.  The negative display uses a square-root stretch.}
\label{OXY}
\end{figure}

\begin{figure}
\caption{WFPC2 image of NE section of NGC 6888 taken through
    the F673N filter, which isolates light from [\ion{S}{2}] $\lambda 
    \lambda 6717,6731$.  The negative display uses a square-root stretch.}
\label{SUL}
\end{figure}

\begin{figure}
\caption{Positions and widths of spatial profiles extracted from
            the WFPC2 data, superimposed on the H$\alpha$ image.}
\label{CUTS}
\end{figure}

\begin{figure}
\caption{a.  Volume emissivities from spatial profile 3.  The
             distance into the clump is measured from the point
             towards the star at which the H$\alpha$ profile
             flattened out.  H$\alpha$ is plotted with a solid
             line; \ion{S}{2}, dotted line; \ion{O}{3}, dashed
             line.  b. Photoionization model for  log~Q$_{0}$
             = 49.3 that best matched the observed profile.
             c. The data (solid) with the model (dotted)
             overplotted.  Distance increases away from the
             central star.}
\label{MODEL}
\end{figure}

\begin{figure}
\caption{Plot of model nitrogen, oxygen, and sulfur abundances as
            a function of the total H-ionizing flux log~$Q_{0}$.
            The lines are identified by their correspondent spatial
            profile number.}
\label{ABUN}
\end{figure}

\begin{figure}
\caption{Spatial profile of the outer shock of NGC 6888.  The
            H$\alpha$ surface brightness has been increased by a
            factor of 5 to ease comparison; [\ion{S}{2}] has been
            raised by a factor of 10.  Distance increases toward the
            central star.}
\label{SHOCK}
\end{figure}

\begin{figure}
\caption{a. Sketch of the wind-blown bubble as the red supergiant (RSG)
            wind is being swept up by the Wolf-Rayet (WR) wind into
            a ``WR shell''.  The ``MS bubble'' is composed of a
            mixture of relic main-sequence  wind material and
            evaporated ``RSG shell'' material.
            b.  The bubble some time after the WR and RSG shells
            collide.  The post-shock WR wind now streams past the
            fragmented RSG shell and directly interacts with the
            MS bubble, driving a shock visible as a ``skin''
            enveloping the nebular shell.  The MS bubble material
            is shadowed from the UV flux of the central star by
            the densest of the knots in the nebular shell, affecting
            its [\ion{O}{3}] flux.}
\label{SKETCH}
\end{figure}

\begin{deluxetable}{ccrrrrr} 
\tablecolumns{7} 
\tablewidth{0pc} 
\tablecaption{Model Parameters for Spatial Profile Fits}
\tablehead{ 
\colhead{}    &  \colhead{}    &  \multicolumn{5}{c}{log Q$_{0}$} \\ 
\cline{3-7} \\ 
\colhead{Spatial Profile} & \colhead{} & \colhead{49.20}   & \colhead{49.25}    
& \colhead{49.30} & \colhead{49.35} & \colhead{49.40}}
\startdata 
1 & Peak n$_{{\rm H}}$(cm$^{-3}$) & 1570 & 1570 & 1540 & 1530 & 1520 \\*
  &  N  &  8.02  &  8.02  &  8.06  &  8.11  &  8.18  \\*
  &  O  &  7.77  &  7.71  &  7.67  &  7.63  &  7.59  \\*
  &  S  &  6.45  &  6.52  &  6.60  &  6.68  &  6.71  \\*
\tableline
2 & Peak n$_{{\rm H}}$(cm$^{-3}$) & 1300 & 1310 & 1320 & 1320 & 1320 \\*
  &  N  &  8.03  &  8.04  &  8.05  &  8.09  &  8.16  \\*
  &  O  &  7.68  &  7.57  &  7.57  &  7.52  &  7.47  \\*
  &  S  &  6.69  &  6.79  &  6.81  &  6.87  &  6.91  \\*
\tableline
3 & Peak n$_{{\rm H}}$(cm$^{-3}$) & 1030 & 1020 & 1010 & 1000 & 1000 \\*
  &  N  &  8.11  &  8.18  &  8.26  &  8.34  &  8.43  \\*
  &  O  &  7.54  &  7.48  &  7.47  &  7.41  &  7.40  \\*
  &  S  &  6.72  &  6.82  &  6.90  &  6.92  &  6.98  \\*
\tableline
4 & Peak n$_{{\rm H}}$(cm$^{-3}$) & 1150 & 1140 & 1140 & 1140 & 1130 \\*
  &  N  &  7.69  &  7.72  &  7.77  &  7.85  &  7.95  \\*
  &  O  &  8.05  &  7.98  &  7.89  &  7.86  &  7.82  \\*
  &  S  &  6.68  &  6.77  &  6.83  &  6.89  &  6.96  \\*
\tableline
5 & Peak n$_{{\rm H}}$(cm$^{-3}$) & 1100 & 1100 & 1100 & 1130 & 1120 \\*
  &  N  &  8.13  &  8.15  &  8.16  &  8.19  &  8.26  \\*
  &  O  &  7.77  &  7.77  &  7.72  &  7.72  &  7.67  \\*
  &  S  &  6.53  &  6.56  &  6.61  &  6.66  &  6.76  \\*
\tableline
6 & Peak n$_{{\rm H}}$(cm$^{-3}$) & 1500 & 1500 & 1490 & 1490 & 1480 \\*
  &  N  &  7.96  &  7.98  &  8.03  &  8.10  &  8.19  \\*
  &  O  &  7.87  &  7.77  &  7.72  &  7.65  &  7.61  \\*
  &  S  &  6.89  &  6.94  &  7.00  &  7.04  &  7.08  \\*
\enddata 
\end{deluxetable} 

\end{document}